\documentclass[preprint,showpacs,preprintnumbers,amsmath,amssymb]{revtex4}

\usepackage{graphicx}
\usepackage{color}
\usepackage{dcolumn}
\usepackage{bm}

\begin{document}

\title{Proposal for the origin of the cosmological constant}

\author{Roberto Dale}
\email{rdale@dfists.ua.es}
\affiliation{%
Departamento de F\'{\i}sica, Ingenier\'{\i}a de Sistemas y Teor\'{\i}a de la Se\~nal,
Universidad de Alicante, 03690, San Vicente del Raspeig, Alicante, Spain    \\
}%

\author{Juan Antonio Morales--Lladosa and Diego S\'aez}
\email{antonio.morales@uv.es; diego.saez@uv.es}
\affiliation{Departamento de Astronom\'{\i}a y Astrof\'{\i}sica, Universidad de Valencia,
46100 Burjassot, Valencia, Spain\\
}%

\date{\today}
%
%
\begin{abstract}

We work in the framework of a simple vector-tensor theory.
The parametrized post-Newtonian approximation of this theory
is identical to that of general relativity.
Our attention is focused on cosmology. In an homogeneous isotropic
universe, it is proved that the
energy density, $\rho_{_{A}}$, of the vector field $A$, and its pressure, $p_{_{A}}$,
do not depend
on time, and also that the equation of state is $\rho_{_{A}} = -p_{_{A}}$.
This means that, in the theory under consideration, there is a cosmological constant,
which is not vacuum energy, but the dark energy of the cosmic
vector field $A$, whose evolution is classical.

\end{abstract}
\pacs{04.50.Kd,95.36.+x,98.80.Jk}
\maketitle
\section{Introduction}
\label{sec1}
We begin with our motivations to study vector-tensor theories. As it
is well known, the analysis of the WMAP (Wilkinson microwave
anisotropy probe) data has pointed out a set of anomalies in the
cosmic microwave background (CMB) angular power spectrum, e.g., the
planar character of the octopole and its alignment with the
quadrupole, \cite{oli04,bie04,sch04,cop06,cop07} the asymmetry
between the north and south ecliptic hemispheres
\cite{han04a,han04b,eri04a,eri04b,eri07}, and so on. These anomalies
strongly suggest that the distribution of CMB temperatures deviates
from statistical isotropy at very large angular scales. Most of
these anomalies have been recently explained \cite{mor08} by using
divergenceless peculiar velocities (superimpositions of the
so-called vector modes) in the framework of general relativity (GR);
nevertheless, in this theory, the angular velocity of these
divergenceless motions decays during the matter dominated era and,
consequently, such motions should be generated close to $z=1100$ to
produce CMB anisotropy after decoupling and before decaying
\cite{mor08,mor07}. This condition seems to be necessary to make
possible the explanation of the low $\ell $ CMB anomalies proposed
in \cite{mor08}; nevertheless, no specific processes are known to
create the required divergenceless velocity fields around $z=1100$.
By this reason, we propose the use of some suitable theory involving
more vector modes than GR. In this alternative theory, the evolution
of the angular velocity could be very different and more appropriate
to explain anomalies. These comments suggest the choice of some
vector-tensor theory, whose four-vector field leads to new vector
modes associated to its spatial part. According to these ideas, we
are performing a study of the vector-tensor theories obtained from
the Lagrangian described in the next section \cite{wil93}. Other
theories will be considered in the future. Our first outcomes on the
evolution of vector modes in vector-tensor theories were presented
in \cite{dal09}.

Before studying the evolution of the vector modes in a given
vector-tensor theory, we should explain some well known
observations; e.g., it should have an appropriate parametrized
post-Newtonian (PPN) limit compatible with solar system
observations. In a first step, we are studying some vector-tensor
theories leading to the same PPN parameters as GR. Moreover, in the
chosen theory, the homogeneous and isotropic cosmological background
should explain, at least, all the observations already explained by
the concordance model. The main goal of this paper is the
description of a theory with these characteristics. We have found
one of them, in which the vector field plays the role of the
cosmological constant. No vacuum energy is then needed to explain
$SN$ $Ia$ observations, CMB anisotropies, baryonic acoustic
oscillations, and so on. These surprising results have motivated
this brief paper.

We think that the new classical origin of the cosmological constant
is a very interesting result. It deserves attention even if CMB
anomalies cannot ever be explained in the framework of the selected
vector-tensor theory. The evolution of the vector modes
(perturbations of the background four-vector used here) is not
required in this paper. It is necessary to try to explain anomalies
(our initial motivation), but this possible explanation is yet to be
studied. Results will be presented elsewhere.

In this paper, $G$, $a$, $\tau $, and $z$ stand for the gravitation
constant, the scale factor, the conformal time, and the redshift,
respectively. Greek (Latin) indexes run from $0$ to $3$ ($1$ to
$3$). Whatever function $\xi$ may be, $\xi^{\star}$ is its present
value and $\dot{\xi}$ ($\xi^\prime$) denotes its partial derivatives
with respect to $\tau$ ($z$). Quantity $\rho_{c} $ is the critical
density, $\rho_{r}$ ($\rho_{m}$) is the radiation (matter) energy
density, and $\rho_{b}=\rho_{r}+\rho_{m}$ is the total background
energy density of the cosmological fluid. Since we work in a flat
universe, the present value of the scale factor $a^{\star} $ is
assumed to be unity and, then $a=(1+z)^{-1}$. Quantity $w$ is the
ratio between pressure $p$ and density $\rho $. Units are defined in
such a way that the speed of light is $c=1$. Finally, indexes are
raised and lowered with the space-time metric.

\section{Vector-tensor theories}
\label{sec2}
In GR, there is a tensor field $g_{\mu \nu}$ which plays the role of
the space-time metric. In the vector-tensor theories, there are two
fields, the metric $g_{\mu \nu}$ and a four-vector $A^{\mu}$.
Several of these theories have been proposed (see \cite{wil93},
\cite{wil06} and references cited there). We have first considered
those based on the action \cite{wil93}:
\begin{equation}
I = \left( {16\pi G} \right)^{ - 1} \int {\left( {R + \tilde{\omega} A_\mu  A^\mu  R
+ \tilde{\eta} R_{\mu \nu }
A^\mu  A^\nu   - \tilde{\varepsilon} F_{\mu \nu } F^{\mu \nu }
+ \tilde{\gamma} \,\nabla _\nu  A_\mu  \nabla ^\nu  A^\mu
} + L_{m} \right)} \,\sqrt { - g} \,d^4 x
\label{1.1}
\end{equation}
where $\tilde{\omega}$, $\tilde{\eta} $, $\tilde{\varepsilon}$, and
$\tilde{\gamma}$ are arbitrary parameters, $R$, $R_{\mu \nu}$, $g$,
and $L_{m}$ are the scalar curvature, the Ricci tensor, the
determinant of the $g_{\mu \nu}$ matrix, and the matter Lagrangian,
respectively. The symbol $\nabla $ stands for the covariant
derivative, and $F_{\mu \nu} = \nabla_{\mu} A_{\nu } - \nabla_{\nu}
A_{\mu }$. In action (\ref{1.1}), it is implicitly assumed that the
coupling between the matter fields and $A_{\mu}$ is negligible. A
more general action is given in \cite{wil06}. It involves a new term
of the form $\lambda (A_{\mu}A^{\mu}+1)$, where $\lambda$ is a
Lagrange multiplier. With the help of this term, the vector
$A^{\mu}$ is constrained to be timelike with the unit norm. From
this last action, the field equations of the so-called
Einstein-Aether constrained theories can be easily obtained. The
applications of these theories to cosmology are discussed, e.g., in
\cite{zun08}. A mass term of the form $m_{_{A}}^{2} A_{\mu}A^{\mu} $
is used in \cite{boh07} to explain cosmic acceleration with a
massive vector field. The same is done in \cite{belt8} for an
unconstrained theory based on action (\ref{1.1}). Recently, other
theories involving vector fields have also been applied to cosmology
(see, e.g., \cite{tar07,mof06}). Fortunately, among all the proposed
vector-tensor theories, we have found a simple unconstrained one,
based on the action (\ref{1.1}), which leads to a new interpretation
of the concordance model in the absence of vacuum energy.

It can be easily proved that, for $\tilde{\omega} =0$, $\tilde{\eta}
= \tilde{\gamma}$, and arbitrary $\tilde{\varepsilon}$, the PPN
parameters of the theory based on the action (\ref{1.1}) are
identical to those of GR  \cite{wil93}, \cite{bel08}. For this
choice of the free parameters we easily get the following field
equations (see \cite{wil93}):
\begin{equation}
\tilde{\eta}\, (\nabla_{\nu}\nabla^{\nu} A_\mu - R_{\mu \nu } A^\nu)
+2\tilde{\varepsilon}\, \nabla^{\nu}F_{\mu \nu}  =0
\label{1.3.1b}
\end{equation}
\begin{equation}
G_{\mu \nu } \equiv
R_{\mu \nu }  - \frac{1} {2} R g_{\mu \nu }  =
8\pi G T_{\mu \nu }  + \tilde{T}_{\mu \nu }^A   \ ,
\label{1.3.2}
\end{equation}
where
$G_{\mu \nu }$ is the Einstein tensor,
$\tilde{T}_{\mu \nu }^A =- \tilde{\eta} [\Theta_{\mu \nu }^{(\tilde{\eta})}
+ \Theta_{\mu \nu }^{(\tilde{\gamma})}]
-\tilde{\varepsilon} \Theta_{\mu \nu }^{(\tilde{\varepsilon})}
$, and the explicit form of $\Theta_{\mu \nu }^{(\tilde{\eta})}$,
$\Theta_{\mu \nu }^{(\tilde{\gamma})}$, and
$\Theta_{\mu \nu }^{(\tilde{\varepsilon})}$
is
given in \cite{wil93}.
Taking into account the identity
\begin{equation}
\nabla_{\alpha} \nabla_{\beta} A^{\alpha}-
\nabla_{\beta} \nabla_{\alpha} A^{\alpha}=
R_{\alpha \beta} A^{\alpha} \ ,
\label{ident}
\end{equation}
Eq. (\ref{1.3.1b}) is easily rewritten in the form:
\begin{equation}
\tilde{\eta} \nabla_{\mu} (\nabla \cdot A)
+(2 \tilde{\varepsilon}-\tilde{\eta})\nabla^{\nu}F_{\mu \nu}  =0 \ ,
\label{1.3.1}
\end{equation}
where $\nabla \cdot A = \nabla_{\alpha} A^{\alpha}$. By using the
identity (\ref{ident}) and Eq. (\ref{1.3.1}), the energy-momentum
tensor of the field $A_{\mu} $; namely, the tensor $T_{\mu \nu
}^A=\tilde{T}_{\mu \nu }^A/8\pi G$ can be written as follows:
\begin{equation}
\begin{array}{lcl}
T_{\mu \nu }^A
              & = & \displaystyle{\Big[-\frac{\eta}{2}}
                     (\nabla \cdot A)^{2}+(2 \varepsilon-\eta)
                     \Big(A^{\alpha}\nabla^{\beta}F_{\alpha \beta} - \frac{1}{4}
                     F_{\alpha \beta}F^{\alpha \beta}\Big)\Big]g_{\mu \nu } \\
              &  &
                     +(2 \varepsilon-\eta) g^{\alpha \beta}
                     \Big(F_{\mu \alpha}F_{\nu \beta} +A_{\mu}\nabla_{\beta}
                     F_{\nu \alpha} + A_{\nu}\nabla_{\beta}F_{\mu \alpha}
                     \big) \ ;
\end{array}
\label{tmunua}
\end{equation}
where $\eta = \tilde{\eta}/ 8 \pi G$, and $\varepsilon =
\tilde{\varepsilon}/ 8 \pi G$. Finally, in the cosmological case,
$T_{\mu \nu }$ is the energy-momentum tensor of the cosmological
fluid, which involves both matter and radiation. It is worthwhile to
notice that the relation
\begin{equation}
\nabla^{\mu} T_{\mu \nu }^A = 0
\label{diver}
\end{equation}
is satisfied (see \cite{wil93}) and, consequently, taking into account
Eq.~(\ref{1.3.2}) and the identity $\nabla^{\mu} G_{\mu \nu } = 0$,
the relation $\nabla^{\mu} T_{\mu \nu } = 0$ is also
satisfied. This means that matter and radiation evolve as in the
standard Friedmann-Robertson-Walker model of GR and,
as it is well known, after the $e^{\pm} $ annihilation, which took place at $z \sim 10^{10}$
(see \cite{kol94}),
the following equations are valid:
$3p_{r}=\rho_{r} = \rho_{r}^{\star} (1+z)^{4}$, $\rho_{m} = \rho_{m}^{\star} (1+z)^{3}$,
and $p_{m}\simeq 0$.

For $\tilde{\omega} = \tilde{\eta} =  \tilde{\gamma} = 0$ and
$\tilde{\varepsilon} \neq 0$, the term $- \tilde{\varepsilon} F_{\mu
\nu } F^{\mu \nu }$ involved in the action (\ref{1.1}) has the same
form as the term appearing in the action of an electromagnetic field
in the absence of sources. In this case, Eq. (\ref{1.3.1}) reduces
to $\nabla^{\nu}F_{\mu \nu}  =0$ and the energy-momentum tensor of
the field $A_{\mu} $ is $T_{\mu \nu }^A = 2 \varepsilon [ g^{\alpha
\beta} F_{\mu \alpha}F_{\nu \beta} - (1/4) g_{\mu \nu }F_{\alpha
\beta}F^{\alpha \beta}]$. As it is well known, the field $A_{\mu} $
is called the potential vector of $F_{\mu \nu}$. The equations of
the theory can be written in terms of the field $F_{\mu \nu} $ and
its derivatives and, consequently, classical theoretical predictions
and observations refer only to the field $F_{\mu \nu}$. Some
indetermination in the potential field $A_{\mu} $ is unavoidable
(the potential vector can be arbitrarily fixed by using appropriate
gauge conditions). Similarly, for other values of the parameters
involved in the action (\ref{1.1}), the vector field $A_{\mu} $ may
also play the role of a potential generating other measurable
physical fields involved in the field equations. The role of vector
$A_{\mu} $ and the indetermination in it depend on the properties of
the Lagrangian (invariance under transformations of $A_{\mu}$). For
example, in the case $\tilde{\omega} =0$, $\tilde{\eta} =
\tilde{\gamma}$, and $2 \tilde{\varepsilon}=\tilde{\eta} \neq 0$,
the energy-momentum tensor reduces to $T_{\mu \nu
}^A=-\frac{\eta}{2} (\nabla \cdot A)^{2}g_{\mu \nu }$ and
Eq.~(\ref{1.3.1}) gives $\nabla_{\mu} (\nabla \cdot A)=0$. According
to this last equation $\nabla \cdot A$ is constant and,
consequently, tensor $T^{^{A}}_{\mu \nu}$ has the same form as the
energy-momentum tensor corresponding to vacuum; namely, one has
$T^{^{A}}_{\mu \nu} = -\rho_{_{A}} g_{\mu \nu}$, where $\rho_{_{A}}
=\frac{\eta}{2} (\nabla \cdot A)^{2}= constant \neq 0$. This means
that the resulting theory is equivalent to GR plus a cosmological
constant. In this theory, $A_{\mu}$ is not a classical field to be
determined either with the field equations or with observations, it
is a field playing the role of a potential vector for the scalar
field $\nabla \cdot A$, which is the only field (apart from the
metric) involved into the field equations of the theory.

In this paper, we pay particular attention on the theory
corresponding to $\tilde{\omega} =0$, $\tilde{\eta} = \tilde{\gamma}
\neq 0$, and $0 \neq 2 \tilde{\varepsilon} \neq  \tilde{\eta}$. For
this choice of the free parameters of the action (\ref{1.1}), the
field equations (see above) are more complicated than in the simple
case $2 \tilde{\varepsilon}=\tilde{\eta}$; for example, according to
Eq.~(\ref{1.3.1}), the scalar $\nabla \cdot A$ is not constant,
excepting some special physical systems as, e.g., a homogeneous and
isotropic background universe. The evolution of the cosmological
background is studied in the next sections.

\section{Cosmology: basic equations and cosmological constant}
\label{sec3}

In this section, we focus our attention on a homogeneous and
isotropic cosmological background. In the flat case, the line
element is
\begin{equation}
ds^{2} = a^{2} ( -d\tau^{2} + \delta_{ij} dx^{i} dx^{j} )
\label{bmetric} \ .
\end{equation}
Moreover, homogeneity
and isotropy require a vector field whose covariant components are $(A_{0}(\tau ), 0, 0, 0)$.
Then, tensor $F_{\mu \nu}$ vanishes and
\begin{equation}
\nabla \cdot A = - \frac{1}{a^{2}} \left(\dot A_0 + 2 \frac{ \dot a }{a } A_0 \right) \ ;
\label{diverc}
\end{equation}
hence, Eq. (\ref{1.3.1}) reduces to the relation $d(\nabla \cdot A)/d\tau =0$, which
can be rewritten in the form:
\begin{equation}
\ddot A_0  + 2 A_0 \left( \frac{ \ddot{a}}{a} -
3 \frac{ \dot a^2 }{a^2 } \right) = 0  \ .
\label{13.6}
\end{equation}
Equation~(\ref{tmunua})
allows us to find the components of $T_{\mu \nu }^A$. By using these components,
Eqs. (\ref{1.3.2}) can be written as follows:
\begin{equation}
3\frac{\dot a^2 }{a^2 } = 8\pi G a^2 (\rho_{b}+\rho_{_{A}})
\label{13.7}
\end{equation}
and
\begin{equation}
\label{13.8}
- 2\frac{\ddot a}{a} + \frac{\dot a^2 }{a^2 } =  8\pi G a^2 (p_{b} + p_{_{A}})  \ ,
\end{equation}
where
\begin{equation}
\label{13.9}
\rho_{_{A}}=-p_{_{A}}=\frac {\eta}{2} (\nabla \cdot A)^{2} = \frac{\eta}{2a^{4}}
\left( \dot A_0
+ 2 \frac{\dot a}{a} A_0  \right)^{2}  \ ;
\end{equation}
hence, $w_{_{A}} = -1$, as it occurs in the case of the cosmological
constant ($w_{\Lambda} = -1$).

Since we have found that ---in the cosmological case--- the
energy-momentum tensor of the field $A$ has the form
\begin{equation}
\label{emt}
T_{\mu \nu }^A = - \rho_{_{A}} g_{\mu \nu } \ ,
\end{equation}
Eq. (\ref{diver}) implies the relation $\dot \rho_{_{A}} =0$; hence,
the energy density $\rho_{_{A}} $ is constant. This is in agreement
with the relations $\rho_{_{A}}=\frac {\eta}{2} (\nabla \cdot
A)^{2}$ and $d(\nabla \cdot A)/d\tau =0$ previously obtained.

Evidently, if the energy density of the evolving field $A$ is
constant and the relation $\rho_{_{A}}=-p_{_{A}}$ is satisfied, we
can state that this field acts as a cosmological constant. The
possible values of $\rho_{_{A}}$ are discussed in the next section.

\section{Solving the basic cosmological equations}
\label{sec4}
In order to integrate Eqs.~(\ref{13.6})--(\ref{13.8}), the following
new variables are appropriate: $\zeta = 1+z $, $y_{1} =
\tau^{\prime} $, $y_{2} = A_{0}$, and $y_{3} = A_{0}^{\prime}$. In
terms of these variables, the energy density is
\begin{equation}
\label{13.27}
\rho_{_{A}}  = \frac{\eta \zeta^{2}}{2y_1^2 }  (y_3 \zeta  - 2 y_2)^{2} \ ,
\end{equation}
and Eq.~(\ref{13.7}) can be rewritten as follows:
\begin{equation}
\label{13.1}
\eta \zeta^{2} (y_3 \zeta  - 2 y_2)^{2}  + 2y_1^2 \zeta^{3}
\left( \zeta \rho_{r}^{\star}  + \rho_{m}^{\star}  \right)
= \frac{3}{{4\pi G}}  \ ,
\end{equation}
whereas Eq.~(\ref{13.8}) leads to
\begin{equation}
\label{13.13.1}
y_1^\prime  =  - \frac{y_1}{6 \zeta } \left\{
4\pi G \zeta^{2} \left[ 2y_1^2 \rho_{r}^{\star} \zeta^2
- 3\eta (y_3 \zeta  - 2 y_2)^{2} \right]
+9 \right\} \ .
\end{equation}
Finally, the second order differential equation (\ref{13.6})
leads the following system of first order equations:
\begin{equation}
\label{13.13.2}
y_2 ^\prime   = y_3
\end{equation}
and
\begin{equation}
\label{13.13.3}
y_3^\prime  = \frac{ 2y_2 y_1  +
y_1^\prime \zeta \left( y_3 \zeta
-2 y_2  \right) } {y_1 \zeta^2 }
\end{equation}

From Eqs.~(\ref{13.1}) and (\ref{13.13.1}) one easily finds
\begin{equation}
\label{14.1}
\frac {y_1^\prime}{y_1^3} =
-\frac {4\pi G}{3} (4\rho_{r}^{\star}\zeta^{3}+3\rho_{m}^{\star}\zeta^{2}) \ ,
\end{equation}
which can be easily integrated to get
\begin{equation}
\label{14.2}
y_1 = - \left[ \frac {3} {8\pi G
( \rho_{r}^{\star}\zeta^{4}+\rho_{m}^{\star}\zeta^{3}
+C )} \right]^{1/2}   \ .
\end{equation}
By combining Eqs. (\ref{13.27}), (\ref{13.1}), and (\ref{14.2}) we
can easily get the equality $C = \rho_{_{_A}} $; therefore, taking
into account that $C $ is an arbitrary integration constant, the
constant $\rho_{_{_A}} $ can take on any value and, in particular,
we can fix the value $\rho_{_{A}}=\rho_{_{A}}^{\star} =
\rho_{c}^{\star} -\rho_{m}^{\star}-\rho_{r}^{\star}$ and
$\rho_{m}^{\star} \simeq 0.27 \rho_{c}^{\star}$; thus, we will have
a flat universe with a cosmological constant whose density parameter
is $\Omega_{_{A}} \simeq 0.73$ and, consequently, we can say that we
have got a model explaining the same observations as the concordance
model. In the new model, the dark energy is not vacuum energy, but
the energy of the $A$ field. Since the $\rho_{_{A}}$ value has been
fixed, and the relation $C=\rho_{_{A}}$ is satisfied,
Eq.~(\ref{14.2}) fully defines the function $y_{1} =y_{1}(\zeta)$,
which can be substituted into Eq.~(\ref{13.13.3}) to solve the
system formed by this equation and Eq.~(\ref{13.13.2}). In order to
find a numerical solution of this system, the initial values of
$y_{2}$ and $y_{3}$ are necessary. We have taken $z_{in}=10^{8} $ to
be well inside the radiation dominated era and after $e^{\pm}$
annihilation. As it is well known, in this era, the scale factor is
proportional to the conformal time; hence, one can write $a = \alpha
\tau$, where $\alpha$ is a constant. Moreover, close to $z_{in}$,
the component $A_{0} $ will be approximately  proportional to some
power of $\tau$, namely, $A_{0}=\beta \tau^{\delta}$. If these power
laws are substituted into Eq.~(\ref{13.6}), we obtain the relation
$\delta^{2} - \delta -6 = 0$, whose solution $\delta^{+} = +3$
($\delta^{-} = -2$) defines a growing (decaying) mode. We then use
the growing mode $A_{0}=\beta^{+} \tau^{3}$ plus the relation $\tau
= (\alpha \zeta)^{-1}$ to get the following initial values for the
variables $y_1$, $y_2$, and $y_3$:
\begin{equation}
\label{ies}
y_{1in}=-\alpha^{-1} \zeta_{in}^{-2}; \,\,\,\,\,\,
y_{2in}=\beta^{+} \alpha^{-3} \zeta_{in}^{-3}; \,\,\,\,\,\,
y_{3in}=-3\beta^{+} \alpha^{-3} \zeta_{in}^{-4} \ .
\end{equation}
Since $\rho_{_{A}} $ is constant, its value can be calculated from the initial
conditions (\ref{ies}) and Eq.~(\ref{13.27}). The resulting formula is
\begin{equation}
\label{initra}
\rho_{A}=\frac {25 \eta (\beta^{+})^2 } {2 \alpha^{4}}
\end{equation}
The initial conditions necessary to solve Eqs.~(\ref{13.13.2}) and (\ref{13.13.3});
namely, quantities $y_{2in}$ and $y_{3in}$ are given by Eqs.~(\ref{ies}), but we need the
explicit values of the constants $\alpha $ and $\beta^{+}$. In order to get these constants,
we proceed as follows: since $y_{1in}$ has been already calculated,
the first of Eqs.~(\ref{ies}) allows us to obtain $\alpha =
- y_{1in}^{-1} \zeta_{in}^{-2} $ and, then, from the known values of $\alpha $
and $\rho_{_{A}}$, plus Eq.~(\ref{initra}), we get
$\beta^{+} = \pm \alpha^{2} (2\rho_{_{A}}/25\eta)^{1/2}$; hence, parameter $\eta $
is arbitrary, but it only can take on positive values.
After fixing $\eta $, there are two $\beta^{+} $
values with the same absolute value and opposite signs. Each of these
values generates a set of initial conditions. Indeed, in Eq.~(\ref{initra}),
it is easily seen that the value of $\rho_{_{A}}$ (estimated to explain the
observations, see above) fixes the product $(\beta^{+})^{2} \eta$,
but the $\eta $ value and the sign of $\beta^{+} $ remain arbitrary.

It is evident that, if functions $y_{2}(\zeta)$ and $y_{3}(\zeta)$
satisfy Eqs.~(\ref{13.13.2}) and (\ref{13.13.3}), and $D$ is an
arbitrary constant, functions $Dy_{2}(\zeta)$ and $Dy_{3}(\zeta)$
also satisfy these equations. Moreover, according to
Eqs.~(\ref{ies}), quantities $y_{2in}$ and $y_{3in}$ are
proportional to $\beta^{+} $, which means that, if we find the
numerical solution of Eqs.~(\ref{13.13.2}) and (\ref{13.13.3}) for
$\beta^{+} = \alpha^{2} (2\rho_{_{A}}/25)^{1/2}$ (this positive
$\beta^{+}$ value corresponds to $\eta=1$), the products of the
resulting functions $y_{2}(\zeta)$ and $y_{3}(\zeta)$ by $D=\pm
1/\eta^{1/2}$ give other solution. We can thus obtain the solution
corresponding to any $\eta $ value and $\beta^{+} $ sign.

\begin{figure}[tbh]
\includegraphics[angle=0,width=0.65\textwidth]{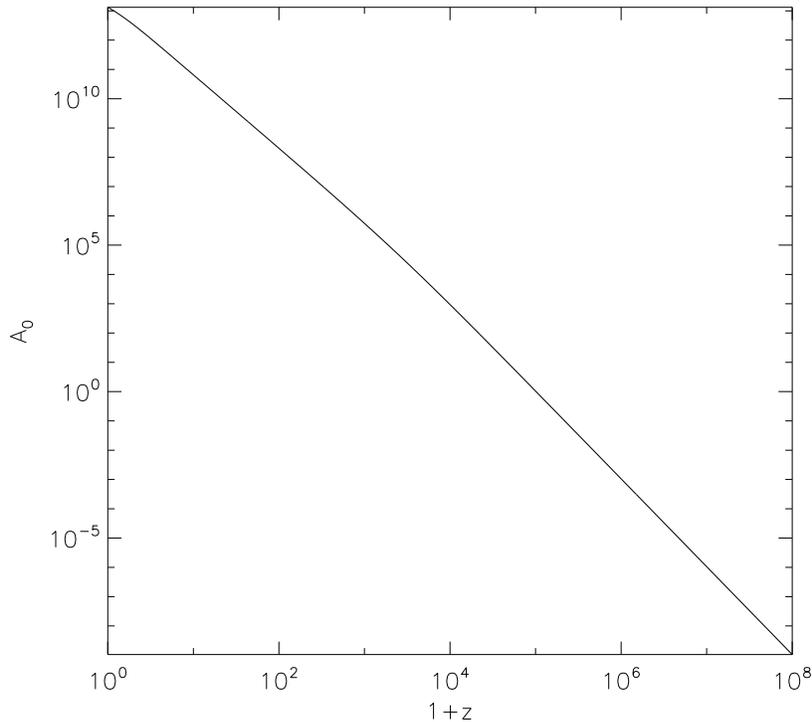}%
\caption{\label{figu1} $A_{0}$ component of the four-vector $A$ as a function of
$1+z$, for $\eta = 1$ and positive $\beta^{+} $.
}
\end{figure}

The $A_{0} $ function corresponding to $\beta^{+} = \alpha^{2}
(2\rho_{_{A}}/25)^{1/2}$ is given in Fig.~\ref{figu1} (in terms of
variable $\zeta $). Equation (\ref{13.1}) can be seen as a
constraint which must be satisfied by functions $y_1$, $y_2$ and
$y_3$. We have verified that the functions we have obtained in our
numerical integration satisfy this equation with a relative error
$(\Xi_1-\Xi_2)/\Xi_2$ smaller than $2 \times 10^{-12}$ for any $z$,
where $\Xi_1$ ($\Xi_2$) is the value of the left- (right-) hand side
of Eq.~(\ref{13.1}). This is a satisfactory numerical test for our
numerical methods and calculations. Very similar results have been
obtained with various numerical methods designed to the integration
of systems of first order differential equations.

\section{General discussion}
\label{sec5}
In the concordance model, we have baryonic matter, dark matter, and
vacuum energy in well-known proportions. The problem with vacuum
energy is that theoretical predictions ---based on standard quantum
field theory--- lead to very big values of this kind of energy,
which are not compatible with those required to explain current
observation (this is the so-called cosmological constant problem).
We might imagine some new quantum theory leading to a strictly
vanishing vacuum energy, but a good enough fitting to the particular
value $0.73 \rho_{c}^{\star} $ seems to be a fine tuning; however,
the vector field $A_{\mu}$ is not normalized either by theoretical
arguments or by non-cosmological observations, which means that, if
this field exists in nature, cosmological data ($\rho_{_{A}}$ value)
may be used, as it is done in previous section, to succeed in the
required normalization. On account of these comments, the new origin
for the cosmological constant seems to be very good news.

In this paper we have proved that a certain cosmic field $A$, with an
appropriate Lagrangian, can play the role of the gravitational constant. This field
is coupled to the scalar curvature and the Lagrangian have the following form:
\begin{equation}
{\cal {L}} = \frac {R}{16\pi G} + \frac{\eta}{2} \left[ R_{\mu \nu } A^\mu A^\nu
\, + \, \nabla _\nu  A_\mu  \nabla ^\nu  A^\mu \right]
- \frac {\varepsilon}{2}F_{\mu \nu}F^{\mu \nu}\ .
\end{equation}
The constant energy density $\rho_{_{A}} $ can take on the right value ($0.73 \rho_{c}^{\star} $)
whatever the values of the $\eta $ and $\varepsilon$ parameters may be. Component $A_{0} $ evolves
(see Fig. \ref{figu1} and Sec.~\ref{sec4}), but $\rho_{_{A}}$ is constant and the
relation $w_{_{A}} = -1$ is valid
at any time. The resulting model is fully equivalent to the standard concordance
model.

The new version of the cosmological constant appears in the study of
the Universe (cosmology). In other cases as, e.g., spherically
symmetric systems, fully asymmetric structures, and so on, the
energy-momentum tensor of the field $A$ must be calculated from the
formula (\ref{tmunua}) of Sec.~\ref{sec2}. We have verified that, in
the static spherically symmetric case where, in adapted coordinates
$(t,r,\theta,\phi )$, the field has the components
$(A_{0}(r),A_{1}(r),0,0)$, the energy-momentum tensor does not
reduce to the form $- \rho_{_{A}} g_{\mu \nu} $. Our cosmological
constant appears in the homogeneous isotropic case; namely, in
cosmology. Also in this sense, we are concerned with a new type of
cosmological constant.

Perturbations of the field $(A_{0},0,0,0)$ are being studied. The
vector part of these perturbations could help to explain the
isotropy violations ---at large angular scales--- detected in the
WMAP maps of CMB temperatures (see Sec. \ref{sec1}). In this way,
the results of papers \cite{mor08,mor07} could be improved. Other
nonlinear applications of the theory could lead to bounds,
relations, or well-defined values of constants $\eta$ and
$\varepsilon$; we are also studying this possibility. Finally,
generalizations of this theory could be studied, e.g., the field
$A^{\mu}$ could be constrained to be timelike with the unit norm,
and a mass term could be introduced in the Lagrangian (see comments
in Sec. \ref{sec2}).

\vspace{2cm}

\begin{acknowledgments}
This work has been supported by the Spanish Ministerio de
Educaci\'on y Ciencia, MEC-FEDER Project No. FIS2006-06062.
\end{acknowledgments}

\vspace{2cm}

\centerline{\LARGE{Addendum}}

\vspace{0.5cm}

While this paper was in production in Phys. Rev. D, we were advised
by A. L. Maroto and J. Beltr\'an Jim\'enez, that similar conclusions
were previously obtained by themselves in arXiv: 0811.0566v1
[astro-ph]. Since the first version of our paper was submitted to
Physical Review on Nov 30 2008 (26 days after the paper 0811.0566v1
was included in the arXiv), the publication in PRD has been
cancelled. In 0811.0566v1, the cosmological constant was interpreted
in the framework of a modified theory of the electromagnetic field
(Einstein-Maxwell generalization), whereas our paper was initially
based on an unconstrained vector-tensor theory which was not
interpreted as a theory of the electromagnetic field. The reader can
verify that both theories seem to be different; however, we have
recently verified --after Beltr\'an\&Maroto complaints-- that they
are based on equivalent Lagrangians whose difference is a total
divergence. Indeed, both papers are based on the general
vector-tensor theory proposed by C. M. Will several decades ago. We
would have never tried any publication based on an explicit modification of
the Einstein-Maxwell equations.

\end{document}